\documentclass{PoS}
\usepackage{graphicx}
\usepackage{epsfig}

\newcommand{\psp}{\psi^{\prime}}

\newcommand{\jpsi}{J/\psi}

\newcommand{\EE}{e^+e^-}

\newcommand{\pp}{\pi^+\pi^-}

\newcommand{\beq}{\begin{equation}}
\newcommand{\eeq}{\end{equation}}
\newcommand{\bitm}{\begin{itemize}}
\newcommand{\eitm}{\end{itemize}}

\newcommand{\ks}{K^0_S}
\newcommand{\mmpp}{\mbox{$M_{\rm miss}$}}
\newcommand{\hbn}{h_b(nP)}
\newcommand{\etabn}{\eta_b(nS)}
\newcommand{\dmhf}{\Delta M_{\rm HF}}

\newcommand{\mevm}{\mathrm{MeV}/c^2}
\newcommand{\fb}{\mathrm{fb}^{-1}}
\newcommand{\Uf}{\Upsilon(5S)}

\newcommand{\Un}{\Upsilon(nS)}

\newcommand{\hb}{h_b(1P)}
\newcommand{\gev}{\,\mathrm{GeV}}
\newcommand{\hbp}{h_b(2P)}
\newcommand{\ee}{e^+e^-}
\newcommand{\mmppsq}{\mbox{$M^2_{\rm miss}$}}
\newcommand{\U}{\Upsilon}
\newcommand{\chibnp}{\mbox{$\chi_{bJ}(nP)$}}
\newcommand{\gevm}{\mathrm{GeV}/c^2}
\newcommand{\pipm}{\pi^{\pm}}
\newcommand{\pimp}{\pi^{\mp}}
\newcommand{\mmpip}{M_{\rm miss}(\pi^+)}
\newcommand{\mmpim}{M_{\rm miss}(\pi^-)}

\newcommand{\mmp}{M_{\rm miss}(\pi)}
\newcommand{\zbo}{Z_b(10610)}
\newcommand{\zbt}{Z_b(10650)}
\newcommand{\mev}{\mathrm{MeV}}

\title{New Heavy Exotic Hadrons}

\ShortTitle{New Heavy Exotic Hadrons}

\author{\speaker{Chengping Shen}\thanks{On behalf of the Belle collaboration and
 supported by  a Grant-in-Aid for Scientific
 Research on Innovative Areas ``Elucidation of New Hadrons with a
 Variety of Flavors'' from the ministry of Education, Culture,
 Sports, Science and Technology of Japan and a Grant-in-Aid for
 for Young Scientists (B) under contract 24740158.}\\
        Graduate School of Science, Nagoya University, Nagoya, Japan\\
        E-mail: \email{shencp@phys.hawaii.edu}}

\abstract{We review recent studies on exotic states at the Belle experiment.
The results include: (1) The measurement of the cross
sections of $\gamma \gamma \to \omega \phi$,
$\phi\phi$, and $\omega \omega$  for masses that range from threshold to 4.0 GeV.
In addition to signals from well
established spin-zero and spin-two
charmonium states, there are clear resonant structures below charmonium threshold,
which have not been previously observed. We report a spin-parity analysis for the new structures;
(2) No $X(3872)$ signal is observed in $\eta J/\psi$ or $\gamma \chi_{c1}$
 mode in $B$ decays.  A narrow peak at 3823.5 MeV/$c^2$ (named as
$\psi_2$) to $\gamma \chi_{c1}$ with a significance of
4.2 standard deviations including systematic uncertainty is
observed in $B^{\pm} \to K^{\pm} \gamma \chi_{c1}$;
(3) The bottomonium states $h_b(1P)$, $h_b(2P)$ and $\Upsilon(1D)$
are observed in the reaction $e^+e^- \to \pi^+ \pi^- + X$;
(4) The observation of two narrow charged structures (named as
$Z_b(10610)$ and $Z_b(10650)$) in the mass spectra of the
$\pi^{\pm}\Upsilon(nS) (n=1,2,3)$ and $\pi^{\pm} h_b(mP)$ $(m=1,2)$ pairs
that are produced in association with a single charged pion in Y(5S) decays.}

\FullConference{Sixth International Conference on Quarks and Nuclear Physics,\\
		April 16-20, 2012\\
		Ecole Polytechnique, Palaiseau, Paris}

\begin{document}

\section{Introduction}

In hadronic physics, the best understood quark-antiquark systems are heavy quarkonia,
i.e, $c\bar{c}$ or $b \bar{b}$ mesons. The discovery of the missing $c \bar{c}$
or $b \bar{b}$ states and the precise measurements of properties of the
already observed ones are important.

The QCD-motivated models predict the existence of hadrons of more complex structure
than conventional mesons or baryons, such as hybrids, multiquark states of either
molecular, tetraquark  or hadrocharmonium configuration. As the conventional
hadron spectrum is much cleaner than the dense spectrum of light states, exotic
states containing $c\bar{c}$ or $b \bar{b}$ are expected  
to be identified more easily than the ones predicted in the
light spectrum. Any resonance observed in addition to predicted multiplets might give a hint of
such an exotic spectroscopy.

Some of the recently observed charmonium-like or bottomonium-like XYZ states could be candidates for the
exotic hadrons mentioned. However most of them still await confirmation or their properties need to be
further studied before any decisive interpretation is made. Here, we review some recent results
on the exotic hadrons from Belle experiment.

\section{Observation of new resonant structures in $\gamma \gamma \to \omega \phi$, $\phi
\phi$ and $\omega \omega$}

Recently a clear signal for a new state
$X(3915)\to \omega \jpsi$~\cite{x3915} and evidence for
another state $X(4350) \to \phi \jpsi$~\cite{x4350}
have been reported, thereby introducing new puzzles to
charmonium or charmonium-like spectroscopy.
It is natural to extend the above theoretical picture to similar
states coupling to $\omega \phi$, $\omega \omega$ or $\phi\phi$.

The measurements of the cross sections for
$\gamma\gamma\to VV$~\cite{gg2vv}, where $VV=\omega\phi, \phi\phi$ and
$\omega\omega$, are based on an analysis of an 870~fb$^{-1}$ data sample taken at or
near the $\Upsilon(nS)$ ($n=1,...,5$) resonances with the Belle detector
operating at the KEKB asymmetric-energy $\EE$ collider.

\begin{figure}[htbp]
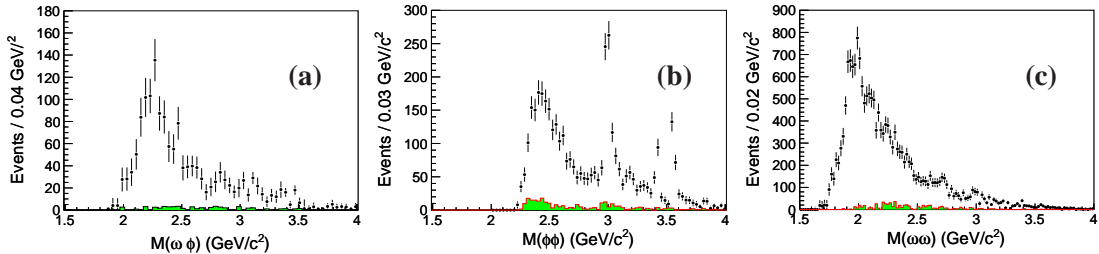

\psfig{file=fig1a.epsi,height=4.7cm, angle=-90}
 \put(-30,-30){ \bf (a)}\hspace{0.1cm}
\psfig{file=fig1b.epsi,height=4.7cm, angle=-90}
 \put(-30,-30){ \bf (b)}\hspace{0.1cm}
\psfig{file=fig1c.epsi,height=4.7cm, angle=-90}
 \put(-30,-30){ \bf (c)}
\caption{ The (a) $\omega \phi$, (b) $\phi \phi$ and (c) $\omega \omega$ invariant mass distributions.
The shaded histograms are from the corresponding
normalized sidebands, which will be subtracted in calculating
the final cross sections.} \label{mass}
\end{figure}

After event selections, clear $\omega$ or $\phi$ signal is observed.
The magnitude of the vector sum of the final particles' transverse
momenta in the $\EE$ center-of-mass (C.M.) frame, $|\sum {\vec
P}_t^{\ast}|$, which approximates the transverse momentum of the
two-photon-collision system, is used as a discriminating variable
to separate signal from background.
We obtain the number of
$VV$ events in each $VV$ invariant mass bin by fitting the $|\sum
{\vec P}_t^{\ast}|$ distribution between zero and 0.9 GeV/$c$.
The resulting $VV$ invariant mass
distributions are shown in Fig.~\ref{mass}, where
there are some obvious structures in the low
$VV$ invariant mass region.

Two-dimensional (2D) angular distributions are
investigated to obtain the $J^P$ of the structures. In the process $\gamma
\gamma \to V V$, five angles are kinematically independent.
We choose $z$, $z^*$, $z^{**}$,
$\phi^*$, and $\phi^{**}$~\cite{angledef} and
use the transversity angle
($\phi_T$) and polar-angle product ($\Pi_\theta$) variables to analyze the
angular distributions. They are defined as \( \phi_T=|\phi^*+
\phi^{**}|/2\pi \), \( \Pi_\theta=[1-(z^*)^2][1-(z^{**})^2] \).

\begin{figure}[htbp]
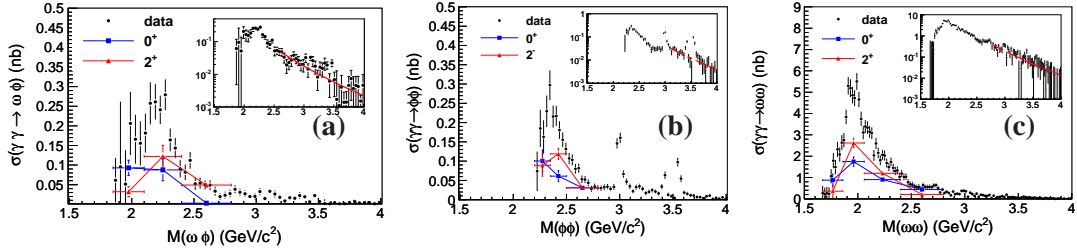

\begin{center}
\includegraphics[height=5.cm,angle=-90]{fig2a.epsi}
 \put(-30,-50){ \bf (a)}\hspace{0.2cm}
\includegraphics[height=4.3cm,angle=-90]{fig2b.epsi}
 \put(-30,-50){ \bf (b)}\hspace{0.2cm}
\includegraphics[height=4.3cm,angle=-90]{fig2c.epsi}
 \put(-30,-50){ \bf (c)}
\caption{The cross sections of $\gamma \gamma \to \omega \phi$
(a), $\phi \phi$ (b), and $\omega \omega$ (c)
are shown as points with error
bars. The cross sections for different $J^P$ values as a
function of $M(VV)$  are shown as the triangles and squares with error bars.
The inset also shows the cross section on
a semi-logarithmic scale. In the high energy region, the solid curve
shows a fit to a $W^{-n}_{\gamma
\gamma}$  dependence  for the cross section after the significant charmonium contributions
($\eta_c$, $\chi_{c0}$ and $\chi_{c2}$) were excluded. }
\label{cross-section}
\end{center}
\end{figure}

We obtain the number of signal events by fitting the $|\sum {\vec P}_t^{\ast}|$
distribution in each $\phi_T$ and $\Pi_\theta$ bin in the 2D
space, which is divided into $4\times 4$, $5\times 5$, and
$10\times 10$ bins for $\omega \phi$, $\phi \phi$, and $\omega
\omega$, respectively, for $M(VV)<2.8$~GeV/$c^2$,
in some wider $VV$ mass bins as shown in Fig.~\ref{cross-section}.
The obtained 2D angular distribution data are fitted with the
signal shapes from MC-simulated samples with different $J^P$
assumptions ($0^+$, $0^-$, $2^+$, $2^-$). We find:
(1) for $\omega \phi$: $0^+$ ($S$-wave) or $2^+$ ($S$-wave)
can describe data with $\chi^2/ndf=1.1$ or 1.2, while
  a mixture of $0^+$ ($S$-wave)
and $2^+$ ($S$-wave) describes data with $\chi^2/ndf=0.9$ ($ndf$
is the number of degrees of freedom); (2) for $\phi \phi$: a mixture
of $0^+$ ($S$-wave) and  $2^-$ ($P$-wave) describes data with
$\chi^2/ndf=1.3$; and (3) for $\omega \omega$: a mixture of
$0^+$ ($S$-wave) and $2^+$ ($S$-wave) describes data with
$\chi^2/ndf=1.3$.

The $\gamma \gamma \to VV$ cross sections are shown in
Fig.~\ref{cross-section}. The cross sections for
different $J^P$ values as a function of $M(VV)$ are also shown in
Fig.~\ref{cross-section}.  While there are substantial
spin-zero components in all three modes, there are also
significant spin-two components, at least in the $\phi\phi$ and $\omega\omega$ modes.

The cross sections for $\gamma \gamma \to \omega \phi$ are
much lower than the prediction of the $q^2 \bar{q}^2$ tetraquark model~\cite{vvreview}
of 1~nb, while the resonant structure in the $\gamma \gamma \to \phi \phi$ mode is
nearly at the predicted position. However, the $\phi \phi$ cross section
is an order of magnitude lower than the expectation in the tetraquark model.
On the other hand, the t-channel factorization model~\cite{alexander} predicted that
the $\phi \phi$ cross sections vary between 0.001~nb and 0.05~nb in the mass region
of 2.0 GeV/$c^2$ to 5.0 GeV/$c^2$, which are much lower than the experimental data.
For $\gamma \gamma \to \omega \omega$, the t-channel factorization model~\cite{alexander}
predicted a broad structure between 1.8 GeV/$c^2$ and 3.0 GeV/$c^2$ with a peak cross section
of 10-30~nb near 2.2 GeV/$c^2$, while the one-pion-exchange model~\cite{AKS} predicted an
enhancement near threshold around 1.6 GeV/$c^2$ with a peak cross section of 13~nb using
a preferred value of the slope parameter.
Both the peak position and the peak height predicted in~\cite{alexander} and~\cite{AKS}
disagree with our measurements.

\section{Charmonium and charmonium-like states}

If $X(3872)$ is a tetraquark state, then it has a $C$-odd parity ($C=-$) partner,
which can dominantly decay into $J/\psi \eta$
and $\chi_{c1}\gamma$. $B^{\pm} \to (J/\psi \eta (\to \gamma \gamma)) K^{\pm}$ and $B^{\pm} \to (\chi_{c1}(\to
J/\psi \gamma) \gamma) K^{\pm}$ decay modes are used in the search for
$C$-odd partner of the $X(3872)$ and other new narrow resonances. In all
the decay modes, $J/\psi$ is reconstructed via $e^+ e^-$ and $\mu^+\mu^-$.
$B$ candidates are identified using energy difference $\Delta E\equiv
E_{B}^* - E_{beam}^*$ and beam-energy constrained mass $M_{\rm bc}\equiv
\sqrt{(E_{beam}^*)^2 - (p_B^{*})^2}$, where $E_{beam}^*$ is the beam
energy in the C.M. frame, and $E_{B}^*$ and $p_B^*$ are the energy and momentum of
the reconstructed particles in the C.M. frame.

\begin{figure}[htbp]
\begin{center}
\includegraphics[angle=0,width=0.50\textwidth]{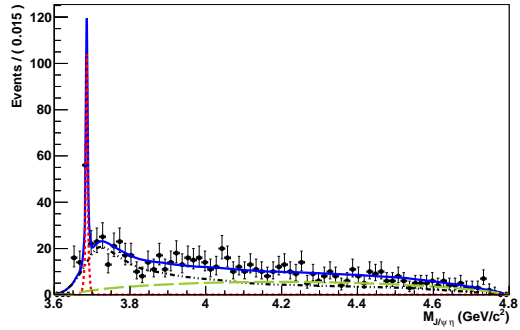}
\caption{The fit to the $M_{J/\psi \eta}$  distribution. Red dashed (green long dashed) curve shows the signal for $B^{\pm} \to \psi' (\to J/\psi \eta) K^{\pm}$ (phase space component $B^{\pm} \to J/\psi \eta K^{\pm}$), while black dashed-dotted curve shows the background parameterized using  $B \to J/\psi X$ MC sample.  }\label{etajpsi}
\end{center}
\end{figure}

The signal region for $B^{\pm} \to J/\psi \eta K^{\pm}$ candidates
is defined as $M_{\rm bc} > 5.27$ GeV$/c^2$ and -35 MeV $< \Delta E< $ 30
MeV. The final $\eta \jpsi$ invariant mass distribution is shown in Fig.~\ref{etajpsi}
together with the fitted results.  No hint of a narrow resonance is
evident from the current statistics. No $X(3872)$ signal is seen and
we obtain the limit $\mathcal{B}(B^{\pm}\to X(3872) K^{\pm})  \mathcal{B}(X(3872)\to J/\psi \eta)<3.8\times 10^{-6}$
at 90\% C.L.

Complimentary to $B^{\pm}\to (J/\psi \eta) K^{\pm}$ study, the search for the
$X(3872)$'s $C$-odd partner is also
carried in $B^{\pm} \to  (\chi_{c1} \gamma) K^{\pm}$ process. Besides the
$X(3872)$'s $C$-odd partner,
we also keep an eye on any other possible narrow charmonium or charmonium-like candidate.

\begin{figure}[htbp]
\begin{center}
\includegraphics[angle=0,width=0.4\textwidth]{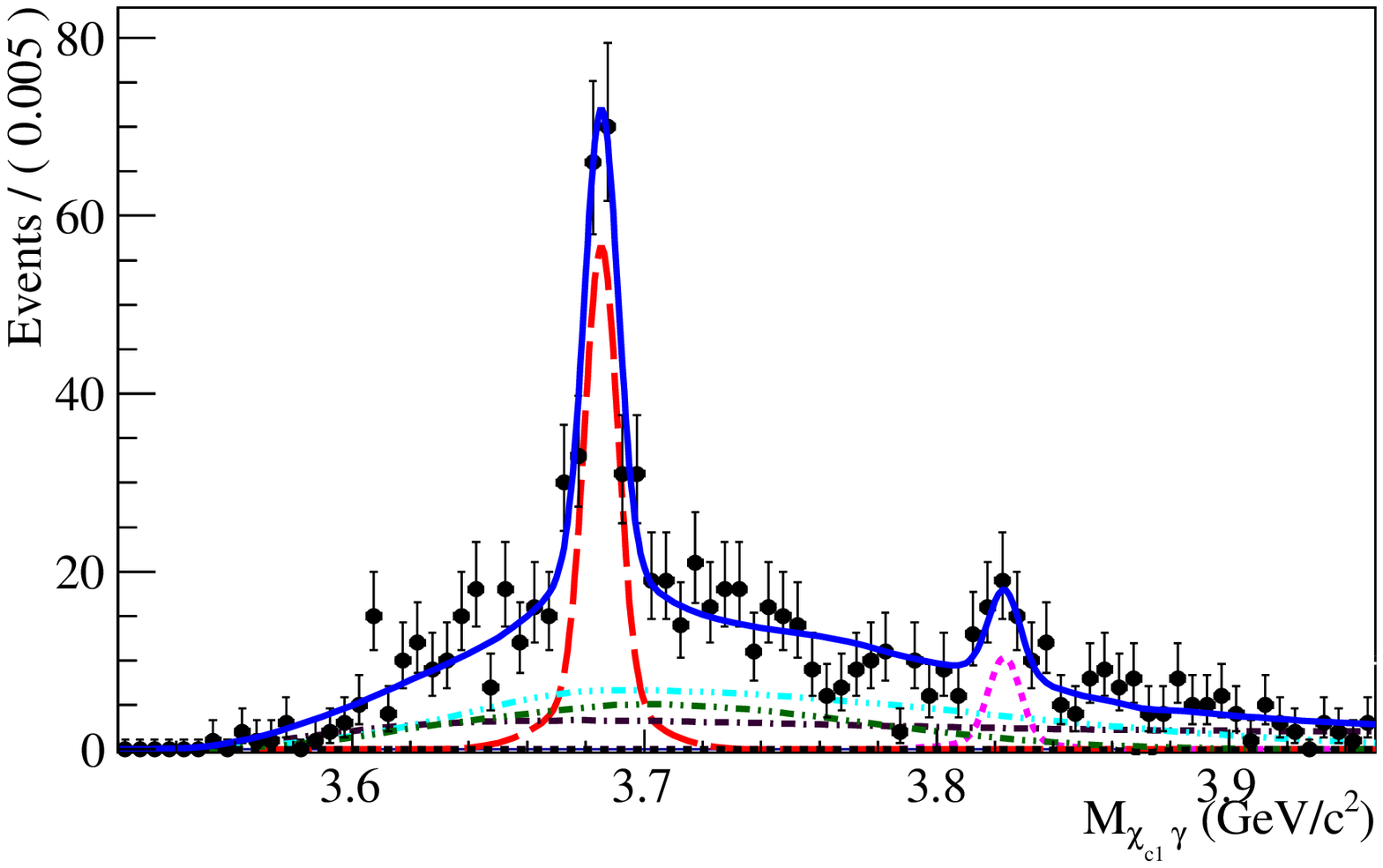}
\includegraphics[angle=0,width=0.4\textwidth]{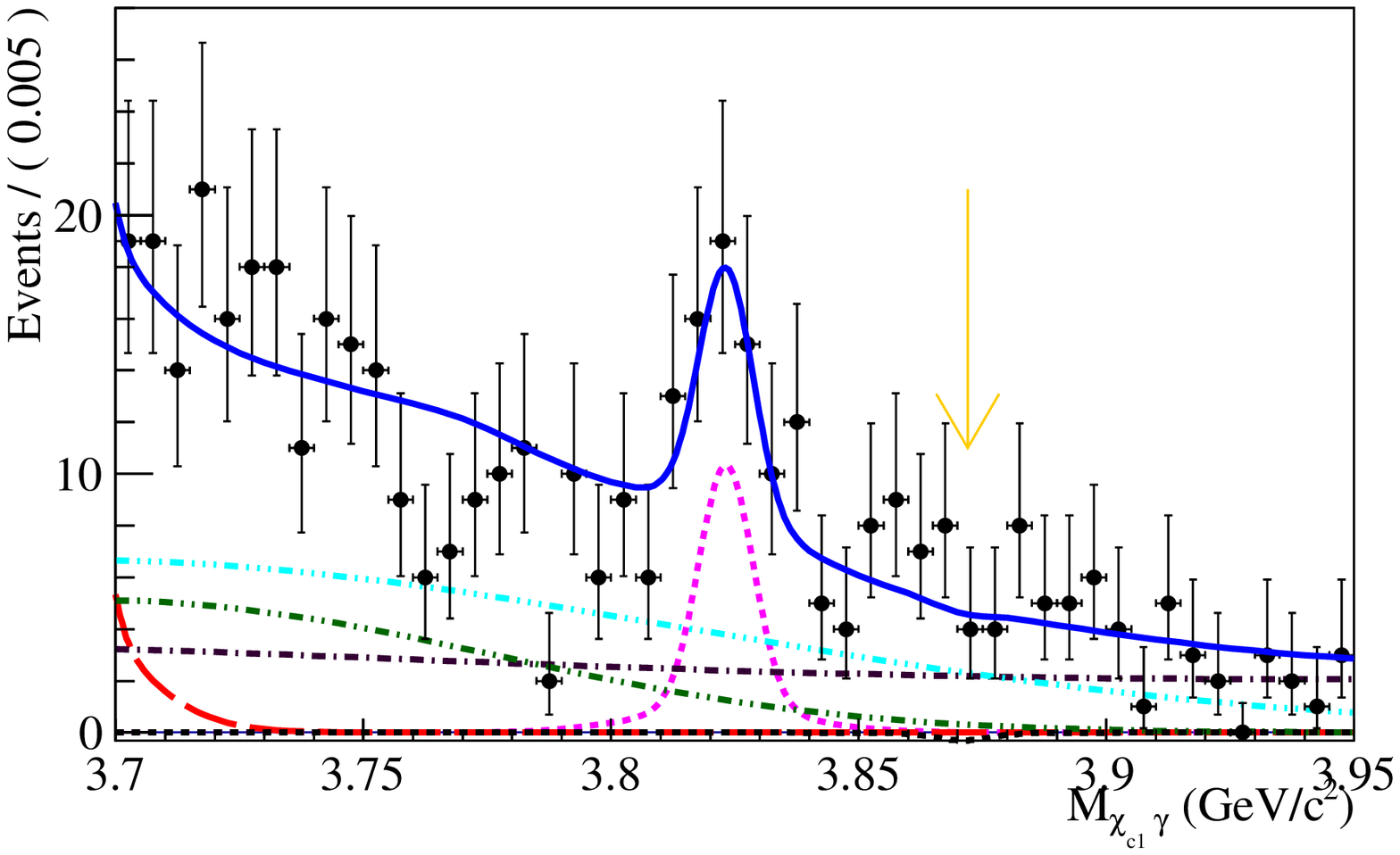}
\includegraphics[angle=0,width=0.4\textwidth]{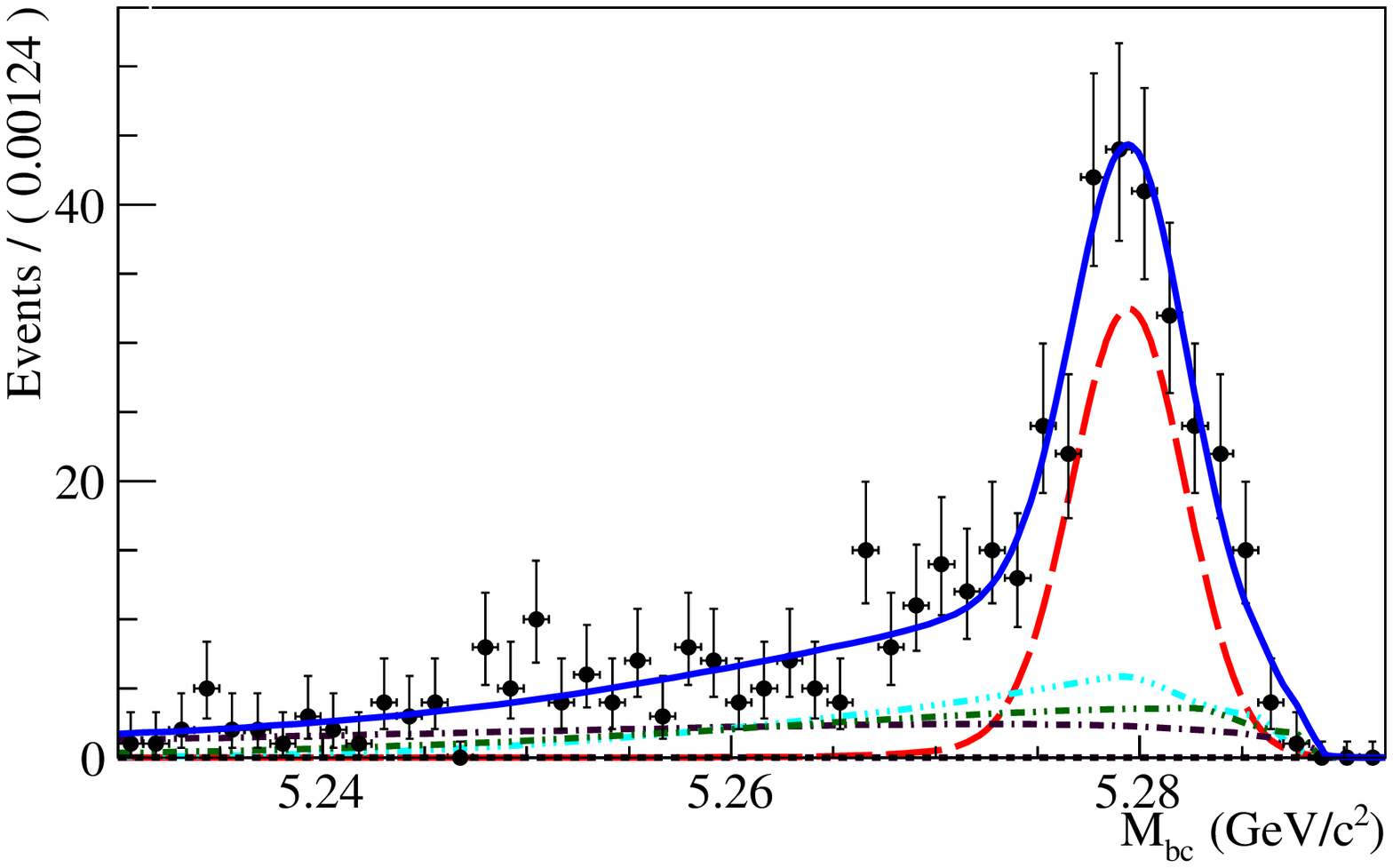}
\includegraphics[angle=0,width=0.4\textwidth]{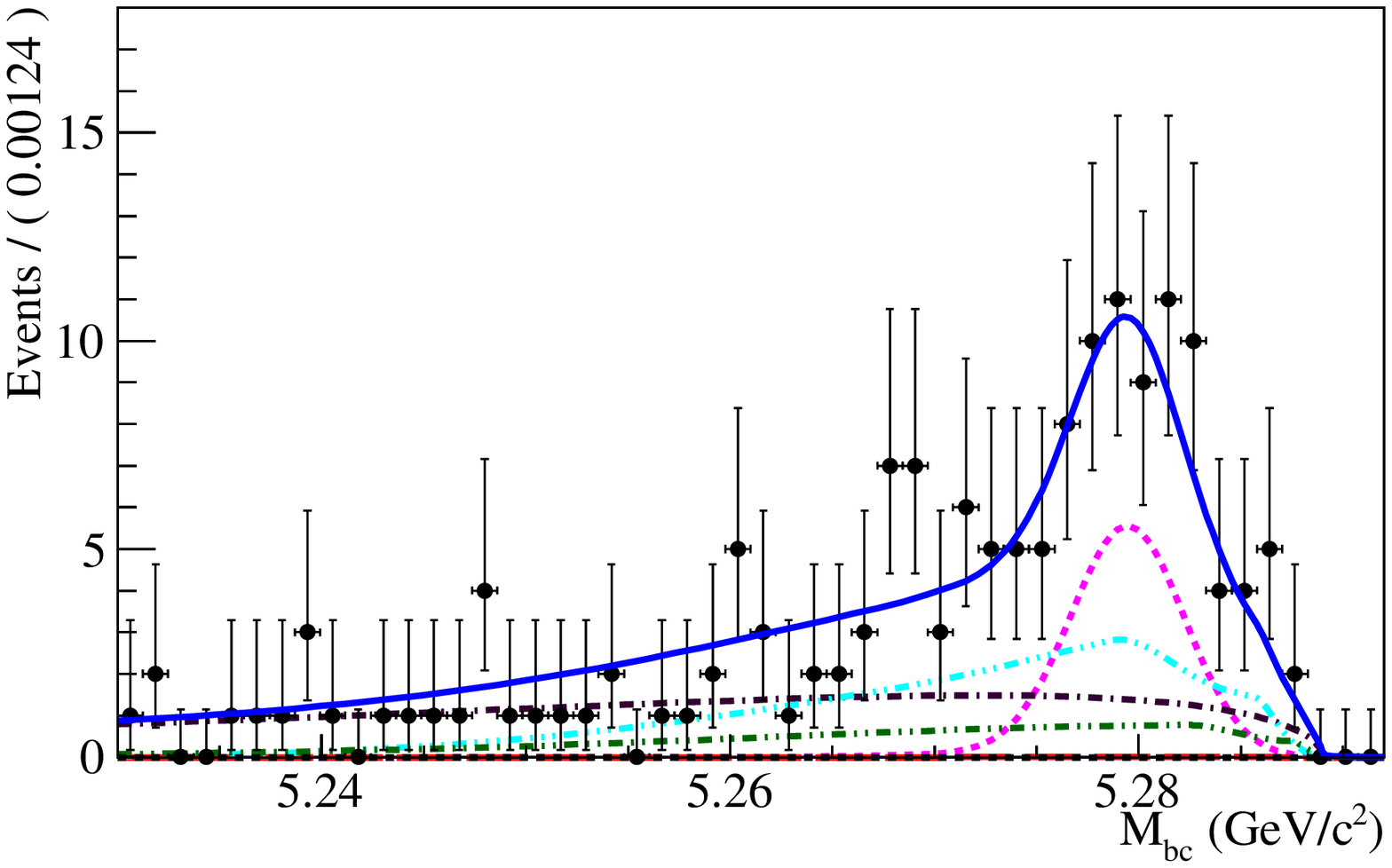}
\caption{The projections from 2D unbinned maximum likelihood fits to
 the $M_{\chi_{c1}\gamma}$ distribution with $M_{\rm bc} > 5.27$ GeV$/c^2$ (top left),
 enlarged $M_{\chi_{c1}\gamma}$ distribution (top right, yellow arrow shows the $X(3872)$ position),
 $M_{\rm bc}$ distribution with $3.66<M_{\chi_{c1}\gamma}<3.708$ GeV$/c^2$ (bottom left),
 and $M_{\rm bc}$ distribution with $3.805< M_{\chi_{c1}\gamma}<3.845$ GeV$/c^2$ (bottom right).
 The dots with error bars are data. The blue solid line is form the overall fit. The
 red large-dashed and pink dashed curves are for $\psi'$ and $\psi_2$ signals, respectively, while
 black dotted-dashed, dark green two dotted-dashed, and cyan three dotted-dashed curves
 are for the combinatorial background, $B^{\pm}\to\psi' ($other than $\chi_{c1}\gamma)K^{\pm}$ events,
 and peaking background component, respectively.}\label{gchic1}
\end{center}
\end{figure}

 After all the event selections, Fig.~\ref{gchic1} shows the $M_{\chi_{c1}\gamma}$ distribution with $M_{\rm bc} > 5.27$ GeV$/c^2$ (top left),
 enlarged $M_{\chi_{c1}\gamma}$ distribution (top right),
 $M_{\rm bc}$ distribution with $3.66<M_{\chi_{c1}\gamma}<3.708$ GeV$/c^2$ (bottom left),
 and $M_{\rm bc}$ distribution with $3.805< M_{\chi_{c1}\gamma}<3.845$ GeV$/c^2$ (bottom right).
The dots with error bars are data, and the blue solid lines are the projections from 2D unbinned
maximum likelihood fits. No $X(3872)$ signal is seen and
the limit $\mathcal{B}(B^{\pm}\to X(3872) K^{\pm})$$\mathcal{B}(X(3872)$$\to $$\gamma \chi_{c1})$$<2.0\times 10^{-6}$
at 90\% C.L. is obtained. Besides the clear $\psp$ signal, we find a clear evidence of a narrow peak
at $3823$ MeV in $M_{\chi_{c1}\gamma}$, as shown in Fig.~\ref{gchic1} with pink dashed curve (named as $X(3823)$).
The signal significance is 4.2 $\sigma$ with systematic error included.
The mass and width of this peak are estimated to be
$3823.5\pm2.1$ MeV$/c^2$ and $4\pm 6$ MeV, respectively. We noticed that
charmonium model predicts a narrow state ($^3D_2$ $c\bar{c}$) at around
3810-3840 MeV/$c^2$~\cite{psi2}. So the $X(3823)$ is probably $\psi_2$ state.
The measured $\mathcal{B}(B^{\pm}\to X(3823) K^{\pm})\mathcal{B}(X(3823)\to\gamma \chi_{c1})$
is $(9.70^{+2.84+1.06}_{-2.52-1.03})\times 10^{-6}$, where the first errors are statistical and the
second systematic.

\section{Bottomonium and bottomonium-like states}

The spin-singlet states $\hbn$ and $\etabn$ alone provide information
concerning the spin-spin (or hyperfine) interaction in bottomonium.
Measurements of the $\hbn$ masses provide
unique access to the $P$-wave hyperfine splitting, $\dmhf\equiv\langle M(n^3P_J)\rangle -M(n^1P_1)$,
the difference between the spin-weighted average mass of the $P$-wave triplet states ($\chi_{bJ}(nP)$ or $n^3P_J$)
and that of the corresponding $\hbn$, or $n^1P_1$.  We use
a $121.4\,\fb$ data sample collected
near the peak of the $\Uf$ resonance ($\sqrt{s}\sim 10.865\gev$) with the Belle detector
to report the first observation of the $\hb$ and
$\hbp$ produced via $\ee\to \hbn\pp$ in the $\Uf$ region~\cite{hbpaper}.

We observe the $\hbn$ states in the $\pp$ missing mass spectrum of hadronic events.
The $\pp$ missing mass is defined as
\(\mmppsq\equiv (P_{\Uf} - P_{\pp})^2,\)
where $P_{\Uf}$ is the 4-momentum of the $\Uf$ determined from the
beam momenta and $P_{\pp}$ is the 4-momentum of the $\pp$
system.  The $\pp$ transitions between $\U(nS)$ states provide
high-statistics reference signals.

\begin{figure}[htbp]
\begin{center}
\hspace*{-.5cm}
\includegraphics[width=0.7\linewidth]{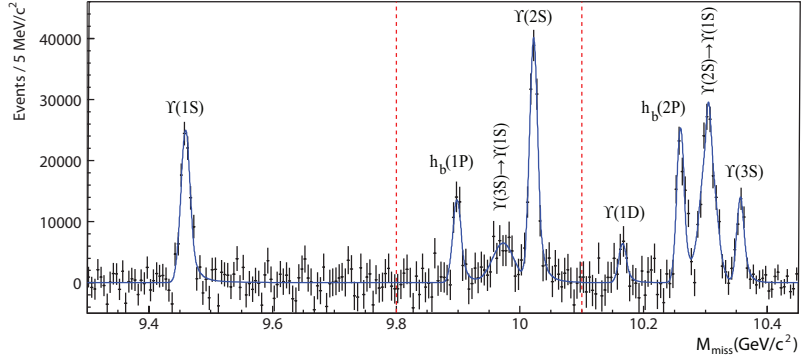}
\caption{ The inclusive $\mmpp$ spectrum with the combinatoric background and
  $\ks$ contribution subtracted (points with errors) and signal
  component of the fit function overlaid (smooth curve).  The vertical
  lines indicate boundaries of the fit regions.}
\label{mmpp_all}
\end{center}
\end{figure}

To reconstruct the $\Upsilon(5S) \to h_b(nP) \pp$ transitions
inclusively, we use a general hadronic event selection.
The $\mmpp$ spectrum after subtraction of both the combinatoric and $\ks\to\pp$ contributions
is shown with the fitted signal functions overlaid in Fig.~\ref{mmpp_all}.
The significances of the $\hb$ and $\hbp$ signals, with systematic uncertainties accounted for,
are $5.5\sigma$ and $11.2\sigma$, respectively. The measured masses of $\hb$ and $\hbp$ are $M=(9898.2^{+1.1+1.0}_{-1.0-1.1})\,\mevm$ and
$M=(10259.8\pm0.6^{+1.4}_{-1.0})\,\mevm$, respectively.   Using
the world average masses of the $\chibnp$ states, we determine the hyperfine splittings to be
$\dmhf=(+1.7\pm1.5)\,\mevm$ and $(+0.5^{+1.6}_{-1.2})\,\mevm$,
respectively, where statistical and systematic uncertainties are combined in quadrature.

\begin{figure}[htbp]
  \centering
  \includegraphics[width=0.3\textwidth]{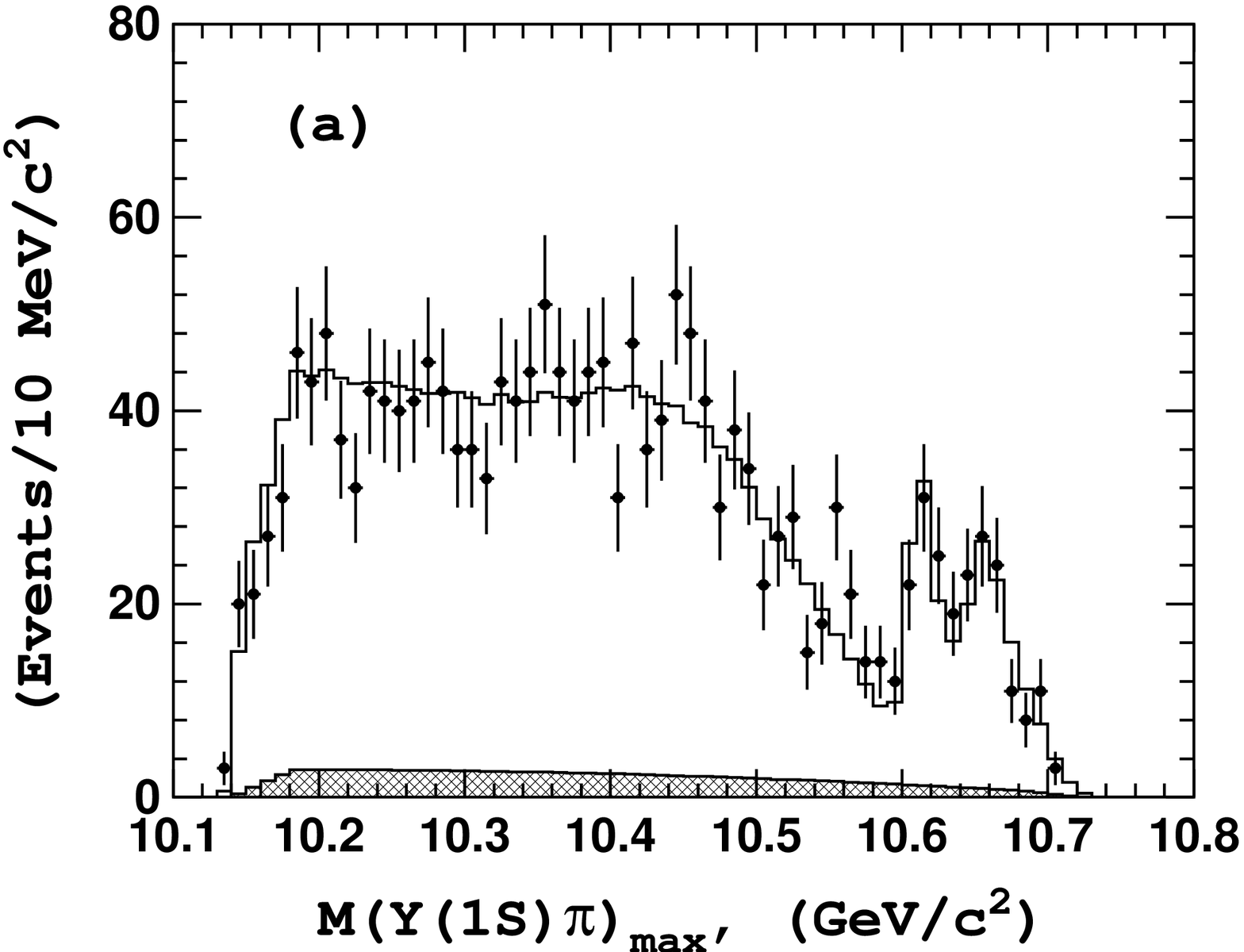}
  \includegraphics[width=0.3\textwidth]{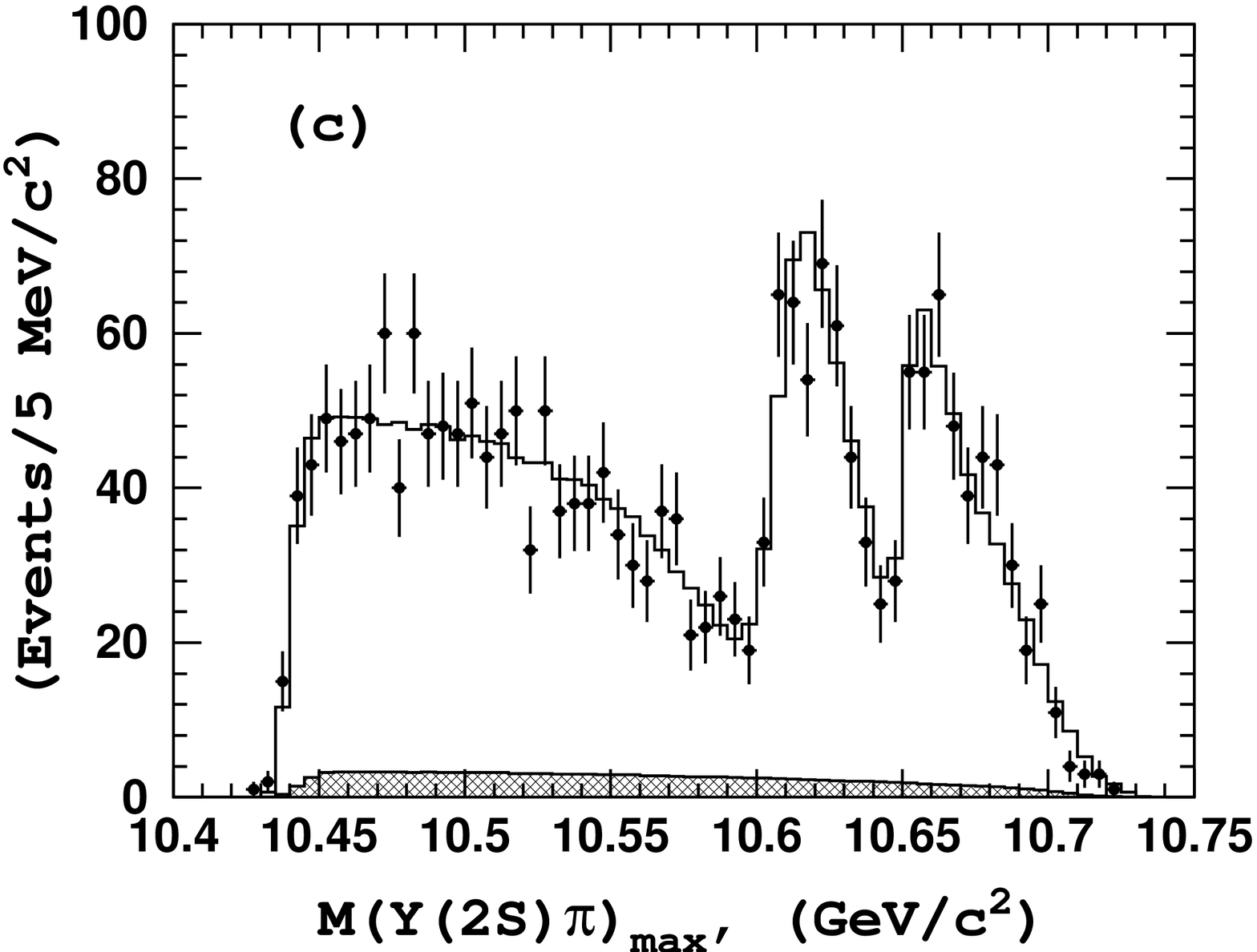}
  \includegraphics[width=0.3\textwidth]{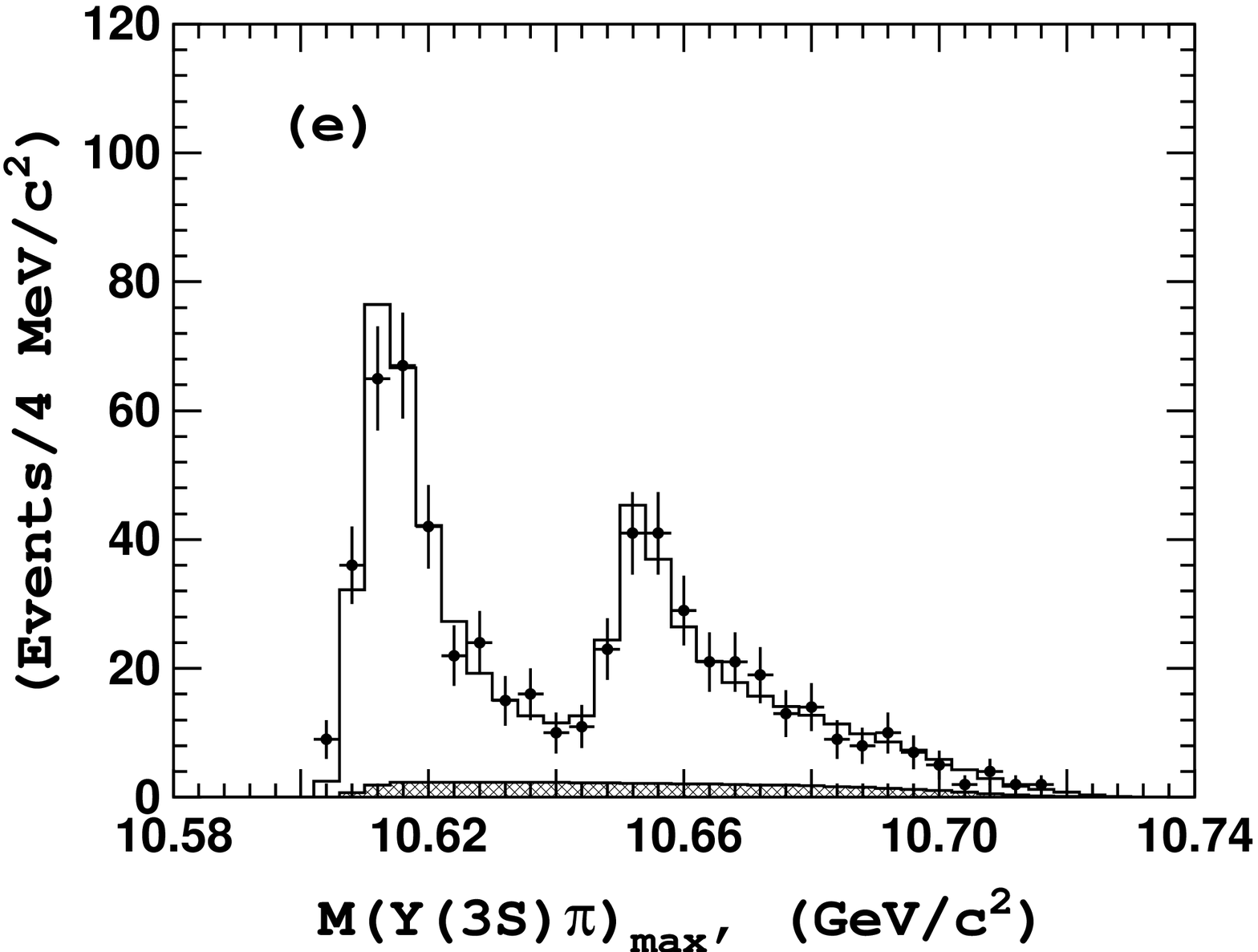}
  \caption{
    Comparison of fit results (open histogram) with
    experimental data (points with error bars) for events in the
    (a) $\Upsilon(1S)$, (c) $\Upsilon(2S)$, and (e) $\Upsilon(3S)$
    signal regions.  The hatched histogram shows the background component.
     }
\label{fig:y3spp-f-hh}
\end{figure}

The observation of anomalously high rates for $\Uf\to\Un\pp$
($n=1,2,3$) and $\Uf\to\hbn\pp$ ($m=1,2$)
transitions suggests  that exotic mechanisms are contributing to $\Uf$ decays.
Amplitude analyses of the three-body $\Uf\to\Un\pp$ decays with $\Un \to \mu^+ \mu^-$ are
performed by means of unbinned maximum likelihood fits to two-dimensional
$M^2[\Un\pi^+]$ vs.\ $M^2[\Un\pi^-]$ Dalitz distributions~\cite{zbb}. One-dimensional invariant mass projections for
events in the $\Un$ signal regions are shown in Fig.~\ref{fig:y3spp-f-hh},
where two peaks are observed in the $\Un\pi$ system near $10.61\,\gevm$
and $10.65\,\gevm$ (named as
$Z_b(10610)$ and $Z_b(10650)$). The combined statistical significance of the two peaks exceeds $10\,\sigma$
for all $\Un\pp$ channels.

To study the resonant substructure of the $\Uf\to\hbn\pp$ ($m=1,2$)
decays we measure their yield as a function of the $\hb\pipm$ invariant
mass. The decays are reconstructed inclusively using the missing mass of
the $\pp$ pair, $\mmpp$. We fit the $\mmpp$ spectra in bins of $\hb\pipm$
invariant mass, defined as the missing mass of the opposite sign pion,
$M_{\rm miss}(\pimp)$. We combine the $\mmpp$ spectra for the corresponding
$\mmpip$ and $\mmpim$ bins and we use half of the available $\mmp$ range
to avoid double counting.

\begin{figure}[htbp]
\vspace*{1mm}
\begin{center}
\includegraphics[width=0.23\textwidth]{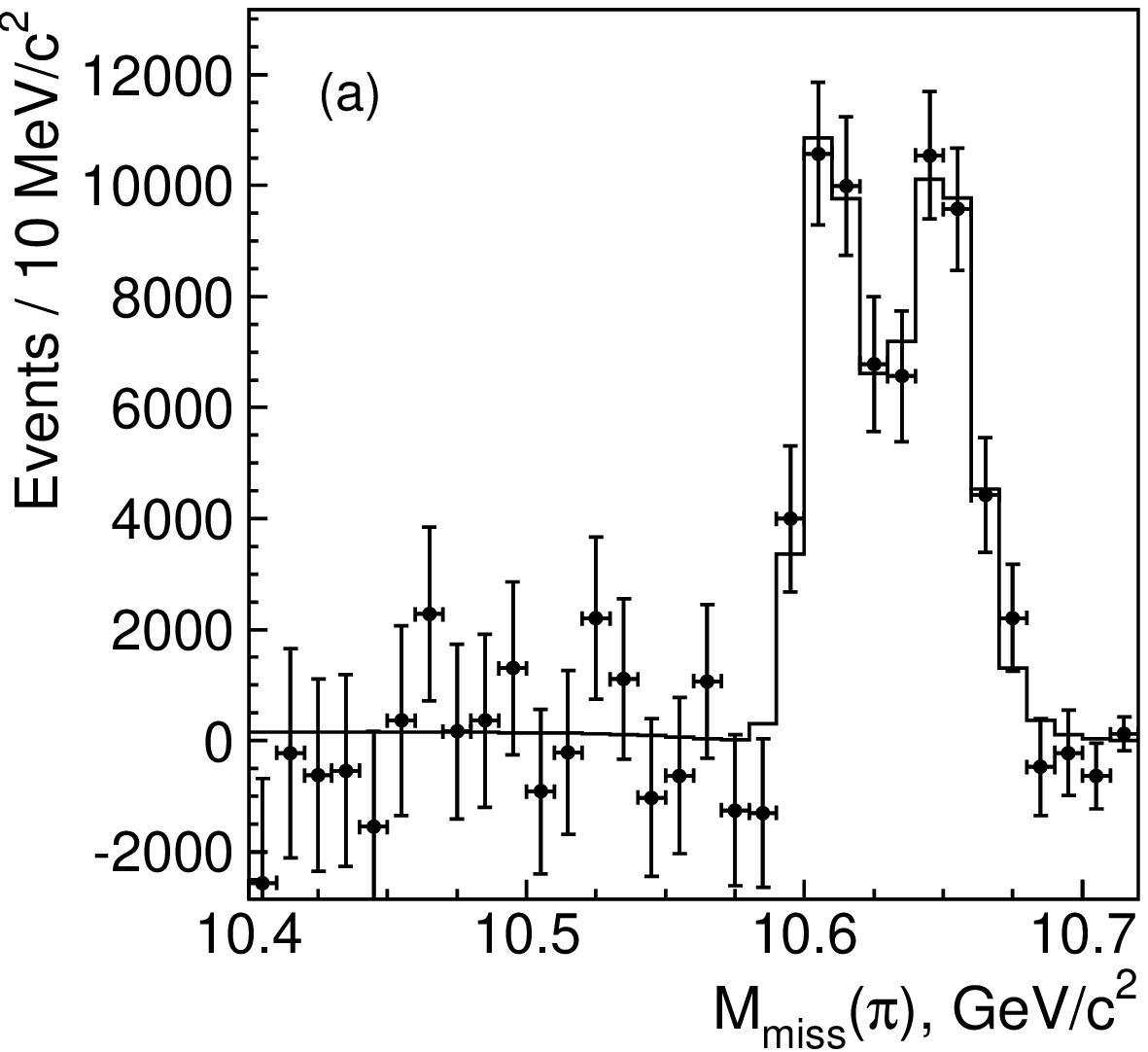}
\includegraphics[width=0.23\textwidth]{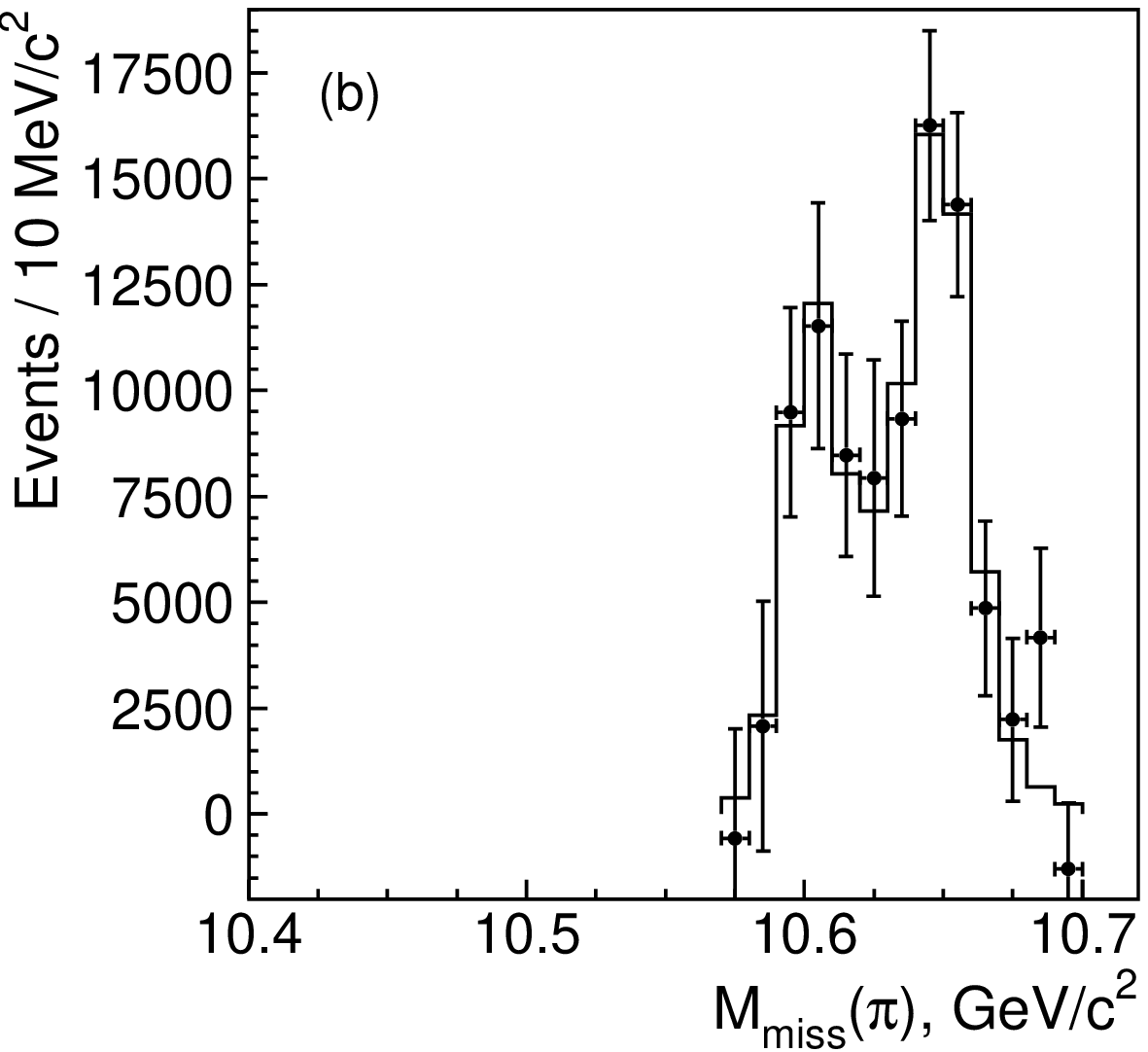}
\caption{The (a) $\hb$ and (b) $\hbp$ yields as a function of $\mmp$
(points with error bars) and results of the fit (histogram).}
\label{fig:mhbpi}
\end{center}
\end{figure}

The results for the yield of $\Uf\to\hbn\pp$ ($m=1,2$) decays as a
function of the $\mmp$ are shown in Fig.~\ref{fig:mhbpi}, where
the fit results are shown as solid
histograms. The two-peak structures are clear in both of them.
The default fit hypothesis is favored
over the phase-space fit hypothesis at the $18\,\sigma$
[$6.7\,\sigma$] level for the $\hb$ [$\hbp$].

Weighted averages over all five channels give
$M=10607.2\pm2.0\,\mevm$,
$\Gamma=18.4\pm2.4\,\mev$ for the $\zbo$ and
$M=10652.2\pm1.5\,\mevm$,
$\Gamma=11.5\pm2.2\,\mev$ for the $\zbt$,
where statistical and systematic errors are added in quadrature.
Angular analysis favors a $J^P=1^+$ assignment for both $Z_b^+$
states, which must also have negative $G$-parity.
Transitions through
$Z_b^+$ to the $h_b(nP)$ saturate the observed $\pp h_b(nP)$ cross sections.
The two masses of $Z_b^+$ states are just a few MeV above
the $B^*\bar{B}$ and $B^*\bar{B^*}$ thresholds, respectively.
The $Z_b^+$ cannot be simple mesons because
they are charged and have $b\bar{b}$ content.

\section {Summary}

We reviewed here some recent results on exotic states at the Belle experiment,
including the measurement of the cross
sections of $\gamma \gamma \to \omega \phi$,
$\phi\phi$, and $\omega \omega$;  the search for the $X(3872)$'s C-odd partner
in $\eta J/\psi$ and $\gamma \chi_{c1}$  modes in $B$ decays;
the evidence of $\psi_2$ in $B^{\pm} \to K^{\pm} \gamma \chi_{c1}$; the observation of
$h_b(1P)$, $h_b(2P)$, $\Upsilon(1D)$, and two charged $Z_b(10610)$ and $Z_b(10650)$ states.


\end{document}